\documentclass[aps,floats]{revtex4}
\usepackage{amsmath}
\usepackage{graphicx,epsfig}
\begin{document}
\bibliographystyle {plain}

\def\oppropto{\mathop{\propto}} 
\def\opsimeq{\mathop{\simeq}}
\def\opoverderline{\mathop{\overline}}
\def\operarrow{\mathop{\longrightarrow}}
\def\opsim{\mathop{\sim}}

\def\fig#1#2{\includegraphics[height=#1]{#2}}
\def\figx#1#2{\includegraphics[width=#1]{#2}}

%\newcommand{\fig}[2]{\epsfxsize=#1\epsfbox{#2}} \reversemarginpar 

%%%%%%%%%%%%%%%%%%%%%%%%%%%%%%%%%%%%%%%%%%%%%%%%%%%%%%%%%%%%%%%%%%%%%%%%%%%%
\title{ On a non-linear Fluctuation Theorem   \\
for the aging dynamics of disordered trap models  } 

\author{C\'ecile Monthus}
\affiliation{Service de Physique Th\'eorique, 
Unit\'e de recherche associ\'ee au CNRS, \\
DSM/CEA Saclay, 91191 Gif-sur-Yvette, France}

%%%%%%%%%%%%%%%%%%%%%%%%%%%%%%%%%%%%%%%%%%%%%%%%%%%%%%%%%%%%%%%%%%%%%%%%%%%%
\begin{abstract}
We consider the dynamics of the disordered trap model, 
which is known to be completely
out-of-equilibrium and to present strong localization effects
in its aging phase. We are interested into the influence
of an external force, when it is applied from the
very beginning at $t=0$, or only after a waiting time $t_w$.
We obtain a ``non-linear Fluctuation Theorem"
for the corresponding one-time and two-time diffusion fronts
 in any given sample,
that implies the following consequences : 
(i) for fixed times, there exists a linear response regime, 
where the Fluctuation-Dissipation Relation
or Einstein relation is valid even in the aging time sector,
in contrast with other aging disordered systems;
(ii) for a fixed waiting time and fixed external field,
the validity of the linear response regime is limited in time
by a characteristic time depending on the external force;
(iii) in the non-linear response regime, there exists a very simple
relation for the asymmetry in the position for
the one-time and the two-times disorder averaged diffusion fronts
in the presence of an external force, 
in contrast to other models of random walks in random media. 
The present non-linear Fluctuation Theorem
is a consequence of a special dynamical symmetry
of the trap model. 
\end{abstract}

\maketitle

\section{Introduction}

The field of non-equilibrium statistical mechanics
is extremely broad, because the non-equilibrium character
of the dynamics may have various origins.
For instance, a system can be out-of-equilibrium
because it is driven in a non-equilibrium steady state
by external boundary conditions or driving forces \cite{derrida,zia,evans}
or because it remains forever non-stationary 
and presents aging, such as in 
coarsening dynamics \cite{coarsening} or in 
disordered models \cite{reviewdisorder,leticia,ritort}.
 
 Although these various classes of systems usually give rise
to different questions and different approaches, 
it seems nevertheless that a central goal in all
these non-equilibrium systems 
has been to study the possible generalizations
of the Fluctuation-Dissipation Theorem, which is an essential result of
Equilibrium statistical mechanics : it relates
 the response to an external perturbation
with the fluctuations existing in the absence of the perturbation.

On one hand, for non-equilibrium steady states,
various generalizations of the 
Fluctuation-Dissipation Theorem have been proposed 
 \cite{ruelle,bertini,hanneyevans},
as well as the Gallavoti-Cohen Fluctuation Theorem \cite{gallavotti}
for deterministic thermostated dynamics under chaoticity assumptions.
This Fluctuation Theorem, which involves 
the time-reversed dynamics, has been then extended to
various stochastic dynamics, such as Langevin dynamics \cite{kurchan},
jump Markov processes and diffusion processes
\cite{lebowitzspohn,depken} and allows to obtain
results in the {\it non-linear } response regime.

On the other hand, in the field of aging dynamics
of disordered models,
the proposed generalization of the Fluctuation-Dissipation Theorem
concerns the linear response regime
and consists in introducing a Fluctuation-Dissipation
ratio to describe the modified relation between the two-time response function
and the two-time correlation \cite{reviewdisorder,leticia,ritort}.
The introduction of this FDT violation ratio for the linear response
regime has been also very useful to characterize the coarsening dynamics
in non-disodered systems \cite{berthier,godrecheluck,mayer}.
Very recently, a non-linear Fluctuation Theorem
has also been proposed for the aging dynamics
of disordered spin models \cite{semerjian},
where however the time-reversal operation used in non-aging systems \cite
{gallavotti,kurchan,lebowitzspohn,depken}
has to be completely redefined.

Another important generalization of the usual Fluctuation-Dissipation Theorem
deals with the the non-linear response of systems
 driven far away from thermal equilibrium,
but where the initial state before the application of an external force
is the thermal equilibrium \cite{jarzynski,crooks}.  
(Note that the initial state in the other approaches
described above is usually not arbitrary but
the Gibbs-Boltzmann distribution in \cite{kurchan} 
or the stationary measure in \cite{lebowitzspohn}).
Remarkably, the Jarzynski non-equilibrium work relation \cite{jarzynski}
has recently become a very useful tool to reconstruct the free-energy
from non-equilibrium single-molecule pulling experiments \cite{manip}.

 This short and very incomplete presentation of recent 
studies of response and fluctuations in non-equilibrium statistical mechanics
already shows that the subject is very rich and very diverse, and that 
some unifying framework is still lacking in comparison
with equilibrium statistical mechanics.

In this paper, we are interested into the response of trap models
on a hypercubic lattice in finite dimension $d$
to an external force field. Trap models provide a simple phenomenological
mechanism for aging \cite{jp92,jpdean,cmjp}.
The one-dimensional version $d=1$ is moreover interesting on its own
since it appears in various physical applications
concerning transport properties in disordered chains
 \cite{alexander,reviewjpag}
or in the dynamics of denaturation bubbles in random DNA sequences \cite{bubbledna}. Its aging and sub-aging
properties in the absence of a bias
have been much studied, either by 
mathematicians \cite{isopi,benarous,reviewbenarous} and by physicists
\cite{bertinjp,c_agingtrap}.  
Recently, the response of the one-dimensional trap model was
studied via scaling arguments and numerical
simulations in \cite{bertinjpreponse}, where various regimes were found
depending on the relative values of the two times considered $(t_w,t)$
and the external applied field $f$, the main results being that 
the Fluctuation-Dissipation Relation (or Einstein relation)
of the linear response regime is valid even in the aging sector,
whereas the response always become non-linear at long times.
The aim of this paper is to 
show that these response properties are actually two direct consequences of  
a non-linear Fluctuation Theorem, that we establish by using a special symmetry
of the master equation of the trap model.
Another important outcome of the non-linear Fluctuation Theorem
is a prediction for the asymmetry induced by the external force
on averaged diffusion fronts.
We will always discuss the $d=1$ case to simplify the presentation,
but it is clear that the results can be straightforwardly generalized
in arbitrary finite dimension $d$. 
However, we will not discuss mean-field trap models,
where previous studies on the Fluctuation-Dissipation 
relation have shown that
the results depend on the observable \cite{fieldingsollich},
and on the choice of the functional form of the rates \cite{ritortmf,sollich}.
For the trap models on lattice in arbitrary finite dimension 
that we consider, there are no such ambiguities
in the choice of observables and hopping rates,
since the natural observable is the position : we 
are interested into the response of 
 the position to an external bias,
and in the thermal fluctuations of the position.

\section{ Stochastic dynamics of the trap model  }

In this Section, we present the Master Equations
that describe the stochastic dynamics of the trap model
in the absence and in the presence of an external field,
and recall some useful notions and properties.

\label{models}

\subsection{ Trap model without external field  }

\subsubsection{Master Equation in a given sample}

The dynamics of the one-dimensional trap model
is defined by the following Master Equation for
the probability $P_t(i)$ to be at time $t$ on site $n$
\cite{reviewjpag}
\begin{eqnarray}
\frac{dP_t(n)}{dt} =  \frac{P_t(n+1)}{2 \tau_{n+1}} + 
\frac{P_t(n-1)}{ 2 \tau_{n-1}} - \frac{P_t(n)}{\tau_n}
\label{masterequation0}
\end{eqnarray}
supplemented by the simple initial condition at $t=0$
\begin{eqnarray}
P_{t=0}(n) = \delta_{n,0}
\end{eqnarray}
This means that after arriving on site $n$,
the particle spends on site $n$ a random time $t$ distributed with the exponential distribution of mean $\tau_n$
\begin{eqnarray}
f_{\tau_n}(t) = \frac{1}{\tau_n} e^{- \displaystyle \frac{t}{\tau_n} }
\label{ftau}
  \end{eqnarray}
and then jumps with equal probability $(1/2)$ to one of its two
nearest-neighbor sites $(n-1)$ and $(n+1)$.

The trapping times $\{\tau_n\}$ are quenched random variables 
in the Master Equation and represent the disorder : 
at each visit to the site $n$, the particle is submitted
to the the same mean
trapping time $\tau_n$.
A given sample corresponds to
a particular realization
of the trapping times on the whole line $\{...,\tau_{-k}, ... ,\tau_0,\tau_1,...,\tau_k,...\}$.

\subsubsection{Law for the disorder}

The trapping times are usually taken to be independent random variables
distributed with a law presenting the algebraic decay \cite{jp92,jpdean}
\begin{eqnarray} 
q(\tau) \oppropto_{\tau \to \infty} \frac{1}{ \tau^{1+\mu}} 
\label{lawtrap}
\end{eqnarray}
This choice comes from the interpretation of 
the trapping times as 
Arrhenius times associated to energies $E_n$
\begin{eqnarray} 
\tau_n=e^{\beta E_n} 
\label{lawtaue}
\end{eqnarray}
where the distribution of the energies $E_n$
presents the exponential tail
\begin{eqnarray} 
\rho(E) \oppropto_{E \to \infty} e^{- \frac{E}{T_g}} 
\label{lawrhoe}
\end{eqnarray}
to mimic the lowest energies statistics
found in the Random Energy Model \cite{rem} and the distribution
of free energy of states 
in the replica theory of spin-glasses \cite{replica}.
This choice is moreover reinforced \cite{jpmezard}
by the exponential tail 
of the Gumbel distribution which represents
one universality classes of extreme-value statistics.
In this interpretation, the exponent $\mu$ in (\ref{lawtrap})
is directly related to the temperature $\mu=T/T_g$ and $T_g$ represents
a glass transition. Indeed for $\mu<1$, the mean trapping time $\int d \tau 
\tau q(\tau)$ is infinite 
and this directly leads to aging effects.
In particular, the characteristic length
scale $\xi(t)$ for the position $n$ at large time $t$
follows the sub-diffusive behavior
obtained by various methods \cite{alexander,machta,reviewjpag,bertinjp,c_agingtrap}
\begin{eqnarray}
\xi(t)  \sim t^{\frac{\mu}{1+\mu}} 
\label{exponentspace} 
\end{eqnarray}
for $\mu<1$ and not the usual scaling in $t^{1/2}$ of the pure random walk.
For more detailed results concerning the localization properties,
the averaged diffusion front as well as aging and sub-aging properties,
we refer the reader to the recent works \cite{bertinjp,c_agingtrap}.

\subsection{ Trap model in the presence of an external field }

In the presence of an external force field $f$,
the master equation (\ref{masterequation0}) 
becomes \cite{reviewjpag}
\begin{eqnarray}
\frac{dP_t^{(f)}(n)}{dt} = 
P_t^{(f)}(n+1) \frac{e^{ - \beta \frac{f}{2} }}{2 \tau_{n+1}} + 
P_t^{(f)}(n-1)\frac{e^{+ \beta \frac{f}{2} }}{ 2 \tau_{n-1}} 
- P_t^{(f)}(n) \frac{e^{+ \beta \frac{f}{2} } +e^{ - \beta \frac{f}{2} } }{2 \tau_{n}}
\label{masterequation}
\end{eqnarray}
where the hopping rates have been chosen 
to satisfy the detailed balance condition
\begin{eqnarray}
e^{-\beta U(n) } W_{ \{n \to n+1 \}}^{(f)}
= e^{ - \beta U_{n+1} } W_{ \{n+1 \to n \}}^{(f)}
\label{detailedbalance}
\end{eqnarray}
where the total energy $U_n$ contains both
 the random energy $(-E_n)$ (\ref{lawtaue}) of the trap $n$
and the potential energy $(-f n)$ linear in the position $n$
induced by the external applied field $f$
\begin{eqnarray}
 U_n = -E_n -f n   
\end{eqnarray}
The condition (\ref{detailedbalance}) guarantees
that in finite sizes, the dynamics will converge
towards the steady state given by Boltzmann equilibrium $P_{eq}(n) \sim e^{- \beta U(n)}$.

\section{ Conservation of the mean position in any given sample }

As already emphasized in \cite{bertinjpreponse}, the trap model
has a very special property : when there is no external field $f=0$
the mean position defined by 
\begin{eqnarray}
 <n>_{f=0}(t)   \equiv  \sum_{n=-\infty}^{+\infty} 
n P_t^{(f=0)}(n)   
\end{eqnarray}
is a conserved quantity in any given sample !
Indeed, the master equation (\ref{masterequation0})
directly yields
\begin{eqnarray}
\frac{ d <n>_{f=0}(t) }{dt}   = 0 
\label{meannconserved} 
\end{eqnarray}
for any realization of the trapping times $\{\tau_n\}$.
This comes from the fact that the hopping rates 
only depend on the initial state, and not on the final state.
As a consequence, when a particle gets out of any trap, it 
always jump with equal probability $(1/2)$ on the right or on the left.
So the trap model possesses a {\it local} symmetry 
for the jump probabilities in the two directions.
This is in contrast with other models of random walks
in random media, such as for instance the Sinai model, 
where this local symmetry 
does not hold, and where as a consequence, there is a non-trivial
mean position in each sample $<n>_{f=0}(t) \neq 0$ \cite{us_sinai},
and it is only after the average over the samples
that the symmetry in the two directions is restored,
i.e. $\overline{ <n>_{f=0}(t) }=0$.

The property (\ref{meannconserved}) is all the more remarkable
that the disorder completely breaks the symmetry $n \to -n$
in any given sample! In particular, the diffusion
front $P_t(n)$ in a given sample is of course not symmetric in $n \to -n$,
and it is only the average over the disordered samples
that is symmetric in $n \to -n$, because to 
each realization $\{ \tau_n \}$,
one may always associate the reversed realization $\{ \tau_{-n} \}$.

\section{ Non-linear Response to a bias applied for $t \geq 0$}

\subsection{ A non-linear Fluctuation Theorem  }

We now consider a given sample of the trap model
and we wish to compare the dynamics in the presence 
of an external field $(-f)$ with respect to
 dynamics in the presence 
of an external field $(+f)$.
It is clear from the two associated master equations (\ref{masterequation})
that we have the remarkable simple property
\begin{eqnarray}
\frac{ P_t^{(+f)}(n) }{ P_t^{(-f)}(n)} = e^{ \beta f n }
\label{ratio+-}
\end{eqnarray}

We may thus express the differences
of the mean position for the two cases as
\begin{eqnarray}
<n>_{ +f}(t) - <n>_{ - f}(t) \equiv
\sum_{n=-\infty}^{+\infty} n \left[ P_t^{(+f)}(n)  - P_t^{(-f)}(n) \right] 
= \sum_{n=-\infty}^{+\infty} n (1-e^{- \beta f n} ) P_t^{(+f)}(n) 
\label{diffmeann}
\end{eqnarray}

\subsection{ Fluctuation-Dissipation Relation in the linear response regime }

A first consequence of (\ref{diffmeann}) is that 
 the susceptibility with respect to the force $f$ reads
\begin{eqnarray}
 \left[ \frac{d <n>_{f}(t) }{df  } \right]_{\vert_{f \to 0} }= \frac{ \beta }{2}
 < n^2>_{f=0} (t)
\label{chit}
\end{eqnarray}
and thus the Fluctuation-Dissipation Relation 
is valid for fixed $t$ in the limit $f \to 0$
 in the form of the Einstein relation
\begin{eqnarray}
 <n>_{f}(t)  \opsimeq_{f \to 0}  \frac{ \beta f }{2}  < n^2>_{f=0} (t)
\label{einsteint}
\end{eqnarray}

\subsection{ Breakdown at long times of the linear response }

A second consequence of (\ref{diffmeann}) is that
for fixed $f$, the linear regime (\ref{einsteint}) of the response
is valid only if the linearization of the factor $(1-e^{- \beta f n} )$
is consistent, and thus it yields the following criterium for
the validity for the linear response
\begin{eqnarray}
\beta f \xi_{(f=0)}(t) <<1
\end{eqnarray}
where $\xi_{(f=0)}(t)$ represents the length scale
 for the displacement at time $t$
in the unbiased trap model
and follows the anomalous scaling (\ref{exponentspace}).
The linear response is thus valid
only up to a characteristic time
presenting the following dependence in the force $f$
\begin{eqnarray}
 t_{\mu}( f) \equiv \left( \frac{1}{\beta f} \right)^{ \frac{1+\mu}{\mu} }
\label{tmuf}
\end{eqnarray}

\subsection{ Consequence for the disorder averaged diffusion front }

A third consequence of (\ref{diffmeann}) is
the following explicit form 
for the symmetry breaking between $(n)>0$ and $(-n)<0$
for the disorder averaged diffusion front in the presence of the external
field $f>0$ 
\begin{eqnarray}
\overline{ P_t^{(f)}(-n) }=  e^{ - \beta f n } 
\overline {P_t^{(f)}(n) }
\label{averageddifffront}
\end{eqnarray}
 To obtain this property, it is convenient to
associate to each realization ${\cal D}\equiv\{ \tau_n \}$ of the disorder 
 the reversed realization ${\cal D}^R \equiv\{ \tau_n^R \}$
simply defined by $\tau_n^R=\tau_{-n}$ for any $n$.
It is then clear that the the dynamics 
in the presence of the external force $(+f)$ in the sample ${\cal D}$
in the coordinate $(n)$ is completely equivalent to the dynamics
in the presence of the external force $(-f)$ in the sample ${\cal D}^R$
in the coordinate $(-n)$, i.e. more explicitly
we have for the diffusion fronts 
\begin{eqnarray}
 P_{\cal D}^{(f)}(n,t) =  P_{{\cal D}^R}^{(-f)}(-n,t) 
\label{reversedsample}
\end{eqnarray}
So the relation (\ref{ratio+-}) between the diffusion
fronts in the same disordered sample in the presence
of two opposite external forces $(+f)$ and $(-f)$,
may also be rewritten as a relation 
between the diffusion
fronts in the two reversed disordered samples in the presence
of the same external force $(+f)$
\begin{eqnarray}
\frac{ P_{\cal D}^{(+f)}(n,t) }{ P_{{\cal D}^R}^{(+f)}(-n,t) } 
 = e^{ \beta f n }  
\end{eqnarray}
and thus the average over the samples ${\cal D}$ immediately yields 
the simple property (\ref{averageddifffront}).
As a comparison with other models of random walks in random media,  
we may cite the Sinai model :
the explicit expressions given in \cite{us_sinai}
for the diffusion front in the presence of a bias
show that the property (\ref{averageddifffront})
is not satisfied for the Sinai model.

\section{ Non-linear Response to a bias applied for $t \geq t_w $}

We now consider the following aging experiment \cite{bertinjpreponse},
which is very usual in the study of disordered systems \cite{reviewdisorder,leticia,ritort} :
the system first evolves with no external force $f=0$
during the interval $[0,t_w]$,
and then evolves in the presence of the external force $(f)$
 for $t \geq t_w$. We consider the two-time diffusion front  
$P^{(f;t_w)}(n,t;n_w,t_w \vert 0,0)$ which represents
the joint probability to be at position $n_w$ at time $t_w$ 
and at position $n$ at time $t$ (where it is assumed that $t >t_w$). 
Since the dynamics takes place with no bias up to $t_w$,
it is clear that for any sample, we have 
from the property (\ref{meannconserved})
\begin{eqnarray}
 <n_w>_{f,t_w} \equiv \sum_{n} \sum_{n_w} n_w 
P^{(f;t_w)}(n,t;n_w,t_w \vert 0,0) =0
\end{eqnarray}
so that the mean position at time $t$ also corresponds
to the mean displacement between $t_w$ and $t$
\begin{eqnarray}
 <n>_{f,t_w}(t) && \equiv \sum_{n} \sum_{n_w} n 
P^{(f;t_w)}(n,t;n_w,t_w \vert 0,0) = <(n-n_w)>_{f,t_w}(t) 
\end{eqnarray}

Again, we wish to compare the dynamics in a given sample
for the two cases $(+f)$ and $(-f)$. 
The generalization of (\ref{ratio+-})
reads
\begin{eqnarray}
\frac{ P^{(+f;t_w)}(n,t;n_w,t_w \vert 0,0) }
{ P^{(-f;t_w)}(n,t;n_w,t_w \vert 0,0) } 
= e^{ \beta f (n-n_w) }
\label{ratio+-t_w}
\end{eqnarray} 
which leads to the following property for the 
relative mean displacement between the two times $t_w$ and $t$
\begin{eqnarray}
<(n-n_w)>_{ +f;t_w}(t) - <(n-n_w)>_{ - f;t_w}(t) \equiv
\sum_{n=-\infty}^{+\infty} \sum_{n_w=-\infty}^{+\infty}
(n-n_w) (1-e^{- \beta f (n-n_w) })
 P^{(+f;t_w)}(i,t;i_w,t_w \vert 0,0) 
\label{diffmeanntw}
\end{eqnarray}

\subsection{ Fluctuation-Dissipation Relation in the linear response regime }

So for fixed times $(t,t_w)$, there is a linear response
at small field $f$ given by
\begin{eqnarray}
 <(n-n_w)>_{f,t_w}(t)  \opsimeq_{f \to 0}  
\frac{ \beta f }{2} < (n-n_w)^2>_{f=0} (t)
\label{einsteinttw}
\end{eqnarray}
which generalizes the one-time Einstein relation (\ref{einsteint}).

\subsection{ Breakdown of the linear response regime }

But for fixed waiting time $t_w$ and fixed external field $f$,
the validity of the linear response is limited in time $t$
by the condition of the validity of the linearization of
the factor $(1-e^{- \beta f (n-n_w) })$ in (\ref{diffmeanntw})
which reads
\begin{eqnarray}
\beta f \xi_{f=0}(t,t_w) <<1
\end{eqnarray}
where $\xi_{f=0}(t,t_w)$ represents the length scale
 for the relative displacement $(n-n_w)$ between the times $t_w$
and $t$ in the unbiased trap model.
In the aging regime where $t_w \to \infty$, $t \to \infty$ with
the ratio $(t-t_w)/t_w>1$ is fixed, the relative displacement scales as
\cite{bertinjpreponse,c_reponsersrg}
\begin{eqnarray}
 \xi_{f=0}(t,t_w) \sim (t-t_w)^{\frac{\mu}{1+\mu}}
\end{eqnarray}
The validity of the linear response regime is thus again limited
in the aging regime by 
\begin{eqnarray}
t-t_w << t_{\mu}( f) 
\end{eqnarray}
in terms of the characteristic time (\ref{tmuf}). 

\subsection{ Consequence for the two-time disorder averaged diffusion front }

The generalization of the relation (\ref{reversedsample}) 
between the diffusion fronts in two reversed disorder samples 
${\cal D}$ and ${\cal D}^R$ to the two-time aging experiment
reads
\begin{eqnarray}
 P_{\cal D}^{(f;t_w)}(n,t;n_w,t_w \vert 0,0) 
=  P_{{\cal D}^R}^{(-f;t_w)}(-n,t;-n_w,t_w \vert 0,0) 
\label{reversedsampletw}
\end{eqnarray}
So the relation (\ref{ratio+-t_w})
between the diffusion
fronts in the same disordered sample in the presence
of opposite external forces $(+f)$ and $(-f)$,
may also be rewritten as a relation 
between the diffusion
fronts in the two reversed disordered samples in the presence
of the same external forces $(+f)$
\begin{eqnarray}
\frac{ P_{\cal D}^{(f;t_w)}(n,t;n_w,t_w \vert 0,0) }
{ P_{{\cal D}^R}^{(f;t_w)}(-n,t;-n_w,t_w \vert 0,0) } 
= e^{ \beta f (n-n_w) }
\end{eqnarray} 
The average over the samples ${\cal D}$ thus yields 
the following generalization of (\ref{averageddifffront})
for the asymmetry in $n$ induced by the external force $f$
on the two-time disorder averaged diffusion front
\begin{eqnarray}
\overline{ P^{(f)}(-n,t;-n_w,t_w \vert 0,0) }=  e^{ - \beta f (n-n_w) } 
\overline {P^{(f)}(n,t;n_w,t_w \vert 0,0) }
\label{averageddifffrontttw}
\end{eqnarray}

\section{ General form of the special symmetry leading to a simple
non-linear Fluctuation Theorem }

In this Section, we investigate the precise condition
 under which a general master equation
will yield a simple non-linear Fluctuation Theorem
such as the one found for the trap model.

\subsection{ Master Equation with detailed balance
in the presence of an external field}

We consider a general Master Equation, where the configuration of
the system is denoted by ${\cal C}$ 
\begin{eqnarray}
\frac{dP_t^{(f)}({\cal C})}{dt} =  
\sum_{ {\cal C}' } P_t^{(f)}({\cal C}') W^{(f)} ( {\cal C}' \to {\cal C} )
- P_t^{(f)}({\cal C}) \sum_{ {\cal C}' }  W^{(f)} ( {\cal C} \to {\cal C}' )
\label{mastergene}
\end{eqnarray}
where the hopping rates in the presence of an external field $f$
linearly coupled to the observable $b$ are chosen as
\begin{eqnarray}
W^{(f)} ( {\cal C} \to {\cal C}' ) = W^{0} ( {\cal C} \to {\cal C}' )
e^{ \frac{\beta f}{2} [ b({\cal C}') - b({\cal C})] }
\end{eqnarray}
to satisfy the detailed balance condition 
\begin{eqnarray}
\frac{ W^{(f)} ( {\cal C} \to {\cal C}' ) }
{ W^{(f)} ( {\cal C}' \to {\cal C} ) }  = 
\frac{ W^{0} ( {\cal C} \to {\cal C}' ) }
{ W^{0} ( {\cal C}' \to {\cal C} ) } 
e^{ \beta f [ b({\cal C}') - b({\cal C})] }
\label{detailedbalancegene}
\end{eqnarray}
We consider the simple initial condition
 \begin{eqnarray}
P_{t=0}^{(f)}({\cal C}) = \delta_{{\cal C},{\cal C}_0 }
\end{eqnarray}
where $({\cal C}_0)$ is some arbitrary given configuration.

\subsection{ Symmetry condition for the total leaving rates }

Let us now define the auxiliary function
\begin{eqnarray}
Q_t ({\cal C}) \equiv  P_t^{(f)}({\cal C})
e^{  \beta f [ b({\cal C}_0) -  b({\cal C}) ] }
\end{eqnarray}
It satisfies the initial condition
\begin{eqnarray}
Q_{t=0} ({\cal C}) \equiv \delta_{{\cal C},{\cal C}_0 } 
\end{eqnarray}
and evolves with the master equation
\begin{eqnarray}
\frac{dQ_t({\cal C})}{dt} = 
\sum_{ {\cal C}' } Q_t({\cal C}') 
 W^{0} ( {\cal C}' \to {\cal C} )
e^{  \frac{\beta f}{2} [ b({\cal C}') - b({\cal C})] }
- Q_t ({\cal C}) 
 \sum_{ {\cal C}' }  W^{0} ( {\cal C} \to {\cal C}' )
e^{ \frac{\beta f}{2} [ b({\cal C}') - b({\cal C})] }  
\label{masterq}
\end{eqnarray}
We wish to compare with the master equation in the presence of the external field $(-f)$ which reads
\begin{eqnarray}
\frac{dP_t^{(-f)}({\cal C})}{dt} =  
\sum_{ {\cal C}' } P_t^{(-f)}({\cal C}') W^{0} ( {\cal C}' \to {\cal C} )
e^{ \frac{\beta f}{2} [ b({\cal C}') - b({\cal C})] }
- P_t^{(-f)}({\cal C}) \sum_{ {\cal C}' }  W^{(0)} ( {\cal C} \to {\cal C}' )
e^{  \frac{\beta f}{2} [ b({\cal C}) - b({\cal C}')] }
\label{mastermoins}
\end{eqnarray}
The ``arriving" hopping rates in the configuration ${\cal C}$ are thus
identical in (\ref{masterq}) and (\ref{mastermoins}), but
the ``leaving" rates out of the configuration ${\cal C}$
are obviously different term by term.
The important quantity is actually  
the total leaving rate
out of a configuration ${\cal C}$ in the presence
of the external field $(+f)$ which reads 
\begin{eqnarray}
w_{out}^{(+f)}( {\cal C} ) \equiv 
 \sum_{ {\cal C}' }  W^{0} ( {\cal C} \to {\cal C}' )
e^{ \frac{\beta f}{2} [ b({\cal C}') - b({\cal C})] }
\label{meanrateoutofc}
\end{eqnarray}
which is of course a priori different from $w_{out}^{(-f)}(\cal C)$.
This means that generically, the local energy landscape around
 the configuration ${\cal C}$ will be distorted in different ways
by the application of $(+f)$ and by the application of $(-f)$.

So the condition to have the simple relation $P_t^{(-f)}=Q_t$, 
leading to the following non-linear Fluctuation Theorem
\begin{eqnarray}
\frac{  P_t^{(f)}({\cal C}) }{ P_t^{(-f)}({\cal C}) }
= e^{  \beta f [ b({\cal C}) -  b({\cal C}_0) ] }
\end{eqnarray}
is the presence of the special symmetry 
\begin{eqnarray}
w_{out}^{(+f)}({\cal C}) = w_{out}^{(-f)}({\cal C}) 
\ \ \ {\rm \ \ for \ any \ configuration \ } \ \ {\cal C}
\label{symcondition}
\end{eqnarray}
This symmetry is satisfied in the trap model
for any configuration in spite of the presence of the disorder,
because the hopping rates of the unbiased case
depend only on the initial configuration
and not on the final configuration
\begin{eqnarray}
  W^{0} ( {\cal C} \to {\cal C}' )
=  W^{0} ( {\cal C}  )
\label{trapcondition}
\end{eqnarray}
and thus the disorder contained in the hopping rates
disappear from the symmetry condition (\ref{symcondition}),
which is then satisfied because the observable $b$ which is the position
can increase or decrease by one with equal probabilities.

\subsection{ Symmetry condition in the linear regime }

At linear order in the external field $f$,
the symmetry condition (\ref{symcondition}) 
simplifies into the following condition
for the unbiased hopping rates 
\begin{eqnarray}
 \sum_{ {\cal C}' }  [ b({\cal C}') - b({\cal C})]
 W^{0} ( {\cal C} \to {\cal C}' ) =0 
\ \ \ {\rm \ \ for \ any \ configuration \ } \ \ {\cal C}
\label{symconditionlinear}
\end{eqnarray}
This exactly means that the mean value $<b>$ is a conserved quantity
for the unbiased dynamics and for an arbitrary initial configuration
\begin{eqnarray}
\frac{d <b>}{dt} \equiv \sum_{ {\cal C} } b({\cal C} )  \frac{dP_t^{(0)}({\cal C})}{dt} =
\sum_{ {\cal C} } P_t^{(0)}({\cal C})
\left[ \sum_{ {\cal C}' } [ b({\cal C}') - b({\cal C})]
  W^{(0)} ( {\cal C} \to {\cal C}' )
 \right]
\label{evolb}
\end{eqnarray}
which vanishes for an arbitrary set of probabilities $P_t^{(0)}({\cal C})$
only if the condition (\ref{symconditionlinear}) is fullfilled.
For the trap model, this condition corresponds to
the conservation of the mean position in any given sample (\ref{meannconserved}).

If the condition (\ref{symconditionlinear}) is satisfied,
then the Fluctuation-Dissipation Relation 
is valid for fixed $t$ in the limit $f \to 0$
 in the form of the Einstein relation
\begin{eqnarray}
 <(b-b_0)>_{ f}(t)  \opsimeq_{f \to 0}  \frac{ \beta f }{2} 
 < (b-b_0)^2>_{f=0} (t)
\label{einsteintforb}
\end{eqnarray}

\subsection{ Discussion }

It would be interesting to find other examples of stochastic dynamics,
which satisfy the dynamical symmetry (\ref{symcondition}),
apart from the trap model considered here. 
A simple way to satisfy the condition (\ref{symcondition})
would be to be able to associate to each transition ${\cal C} \to {\cal C'}$
an other transition ${\cal C} \to {\cal C''}$
which has the same transition rate in zero-field
$W^{(0)} ( {\cal C} \to {\cal C}'' )
=W^{(0)} ( {\cal C} \to {\cal C}' )$ and such that the variation
of the observable $b$ is exactly the opposite 
$b({\cal C}'') -b({\cal C} )
= -( b ( {\cal C}') - b({\cal C} ))$.
It is however a very strong property, since it should be true
for any starting configuration ${\cal C}$,
and we have not been able to find another example
in the most standard dynamics usually considered in 
statistical physics.

\section{ Conclusion }

In this paper, we have obtained that the stochastic
dynamics of trap models has a very special symmetry
that implies a simple non-linear Fluctuation Theorem
for the one-time and two-time diffusion fronts
in any given sample.
We have obtained the following consequences : 
(i) in the linear response regime, the Fluctuation-Dissipation Relation
or Einstein relation (\ref{einsteinttw}) is valid even in the aging time
sector;
(ii) for a fixed waiting time $t_w$ and fixed external field $f$,
the validity of the linear response regime is limited in the time $(t-t_w)$  
by the characteristic time $t_{\mu}(f) \sim (\beta f)^{- \frac{\mu}{1+\mu}}$;
(iii) in the non-linear regime, we have obtained very simple
relations for the asymmetry in the position $n$ induced by an external force
on the one-time and the two-time disorder averaged diffusion fronts. 

The results (i) and (ii) are in full agreement with 
the scaling analysis and the numerical simulations presented in 
\cite{bertinjpreponse}, and we hope that our approach in some sense
``explains'' why aging trap models, which are known to 
present strong localization effects and to be completely
out-of-equilibrium  (see for instance the effective dynamics
described in \cite{c_agingtrap}), nevertheless satisfies the
Fluctuation-Dissipation Relation in the whole linear response regime,
in contrast with many disordered systems which present
a non-trivial violation in the linear response regime \cite{reviewdisorder,leticia,ritort}.

In a future work, we will consider again the same problem
of the effect of an external force on the trap model
with a complementary point of view :
here we have used a special symmetry of the Master Equation
to derive general results 
which are thus valid
in the whole aging phase $0 <\mu<1$ (\ref{lawtrap}) ;
in the forthcoming paper \cite{c_reponsersrg},
we will obtain various exact explicit results
characterizing response properties
 in the high disorder limit $\mu \to 0$, by 
generalizing the previous real-space renormalization approach 
for the unbiased trap model \cite{c_agingtrap}
to include the presence of an external bias.
In particular, we will show that 
the explicit expressions for the averaged diffusion fronts
 indeed satisfy the constraints (iii) on the asymmetry in $n$
obtained here by a general argument.

\begin{acknowledgments}

 It is a pleasure to thank G. Biroli, T. Garel  
and K. Mallick for drawing my attention to useful references.

\end{acknowledgments}

\end{document}